\documentclass{article}
\usepackage{ijcai22}

\usepackage{times}
\usepackage{soul}
\usepackage{url}
\usepackage[hidelinks]{hyperref}
\usepackage[utf8]{inputenc}
\usepackage[small]{caption}
\usepackage{graphicx}
\usepackage{amsmath}
\usepackage{amsthm}
\usepackage{booktabs}
\usepackage{algorithm}
\usepackage{algorithmic}
\usepackage{subfigure,bm,amsfonts,xcolor,wrapfig}
\newcounter{rcounter}

\title{Perceptual Learned Video Compression with Recurrent Conditional GAN}

\author{
Ren Yang$^1$
\and
Radu Timofte$^{1,2}$\and
Luc Van Gool$^{1,3}$
\affiliations
$^1$ETH Z\"urich, Switzerland \\
$^2$Julius Maximilian University of W\"urzburg, Germany \quad
$^3$KU Leuven, Belgium\\
\emails
ren.yang@vision.ee.ethz.ch,
radu.timofte@uni-wuerzburg.de, vangool@vision.ee.ethz.ch
}

\newcommand*{\eg}{\textit{e.g.}}
\newcommand*{\ie}{\textit{i.e.}}
\newcommand*{\etal}{\textit{et al.}}
\newcommand*{\etc}{\textit{etc}}

\begin{document}

\maketitle

\begin{abstract}
  This paper proposes a Perceptual Learned Video Compression (PLVC) approach with recurrent conditional GAN. We employ the recurrent auto-encoder-based compression network as the generator, and most importantly, we propose a recurrent conditional discriminator, which judges on raw vs. compressed video conditioned on both spatial and temporal features, including the latent representation, temporal motion and hidden states in recurrent cells. This way, the adversarial training pushes the generated video to be not only spatially photo-realistic but also temporally consistent with the groundtruth and coherent among video frames. The experimental results show that the learned PLVC model compresses video with good perceptual quality at low bit-rate, and that it outperforms the official HEVC test model (HM 16.20) and the existing learned video compression approaches for several perceptual quality metrics and user studies.
  The codes will be released at the project page: \url{https://github.com/RenYang-home/PLVC}.
\end{abstract}

\section{Introduction}

%

Recent years have witnessed the increasing popularity of end-to-end learning-based image and video compression. Most existing models are optimized towards distortion, \ie, PSNR and MS-SSIM. Recently, Generative Adversarial Networks (GANs) have been used in image compression, to boost perceptual quality \cite{agustsson2019generative,mentzer2020high}.
However, efficient perceptual \emph{video} compression still remains limited.


\begin{figure}[!t]
\centering
\includegraphics[width=1\linewidth]{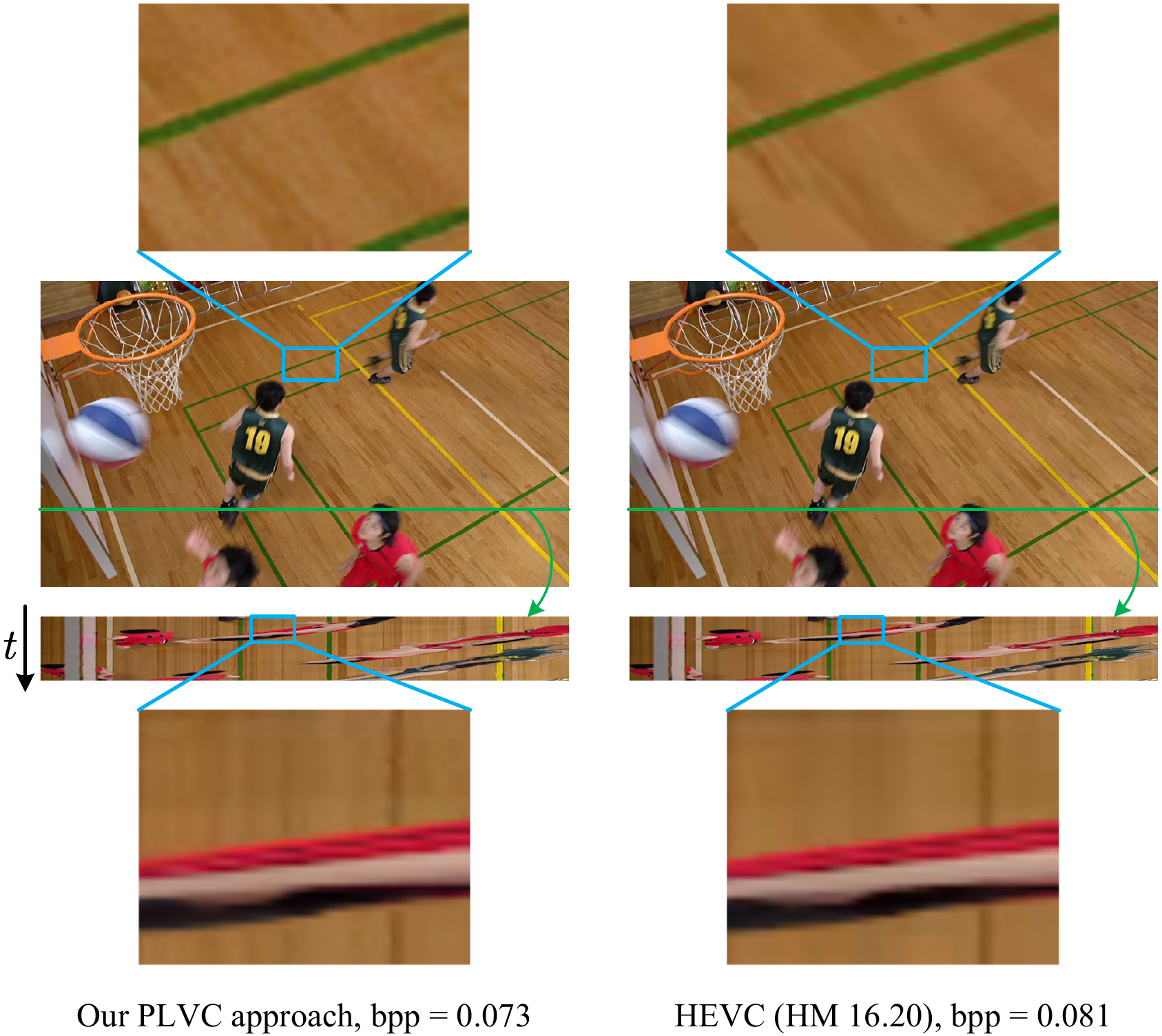}
\caption{Example of the proposed PLVC approach in comparison with the official HEVC test model (HM 16.20).} \label{fig:eg}
\end{figure}

In this paper, we propose a Perceptual Learned Video Compression (PLVC) approach with recurrent conditional GAN, in which a recurrent generator and a recurrent conditional discriminator are trained in an adversarial way for perceptual compression. The recurrent generator contains recurrent auto-encoders to compress video and generate visually pleasing reconstructions in the adversarial training. More importantly, we propose a recurrent conditional discriminator, which judges between raw and compression video conditioned on the spatial-temporal features, including latent representations, motion and the hidden states transferred through recurrent cells. Therefore, in the adversarial training, the discriminator is able to force the recurrent generator to reconstruct both photo-realistic and temporally coherent video, thus achieving good perceptual quality.

Fig.~\ref{fig:eg} shows the visual result of the proposed PLVC approach on \textit{BasketballDrill} (bpp = 0.0730) in comparison with the official HEVC test model HM 16.20  (bpp = 0.0813). The top of Fig.~\ref{fig:eg} shows that our PLVC approach achieves richer and more photo-realistic textures than HEVC. At the bottom of Fig.~\ref{fig:eg}, we show the temporal profiles by vertically stacking a specific row (marked as green) along time steps. It can be seen that the result of our PLVC approach exhibits a temporal coherence comparable with that of HEVC, while offering more detailed textures. As a result, we outperform HEVC on perceptual quality at lower bit-rates.
The contributions of this paper are summarized as:
\begin{itemize}
   \item
   We propose a novel perceptual video compression approach with recurrent conditional GAN, which learns to compress video and generate photo-realistic and temporally coherent compressed frames.
   \item 
   We propose adversarial loss functions for perceptual video compression to balance bit-rate, distortion and perceptual quality.
   \item 
   The experiments (including user study) show the outstanding perceptual performance of our PLVC approach in comparison with the latest learned and traditional video compression approaches.
\end{itemize}

\section{Related work}

\textbf{Learned image compression.} In the past years, learned image compression has gained in popularity. For instance, Ball{\'e}~\etal\ proposed utilizing the variational auto-encoder for image compression and proposed the factorized~\cite{balle2017end} and hyperprior~\cite{balle2018variational} entropy models. Later, the auto-regressive entropy models~\cite{minnen2018joint,lee2019context,cheng2019learning} were proposed to improve the compression efficiency. Recently, the coarse-to-fine model~\cite{Hu2020Coarse} and the wavelet-like deep auto-encoder~\cite{ma2020end} were designed to further advance the rate-distortion performance. Besides, there are also the methods with RNN-based auto-encoder~\cite{toderici2017full,johnston2018improved} or conditional auto-encoder~\cite{choi2019variable} for variable rate compression. Moreover, Agustsson~\etal~\shortcite{agustsson2019generative} and Menzter~\etal~\shortcite{mentzer2020high} proposed applying GANs for perceptual image compression to achieve photo-realistic compressed images at low bit-rate.

\textbf{Learned video compression.} Inspired by the success of learned image compression, the first end-to-end learned video compression method DVC~\cite{lu2019dvc} was proposed in 2019.
Later, M-LVC~\cite{lin2020m} extended the range of reference frames and beats the DVC baseline. Meanwhile, a plenty of learned video compression methods with bi-directional prediction~\cite{djelouah2019neural}, one-stage flow~\cite{liu2019learned}, hierarchical layers~\cite{yang2020learning} and scale-space flow~\cite{agustsson2020scale} were proposed. Besides, methods like the content adaptive model~\cite{lu2020content},  resolution-adaptive flow coding~\cite{hu2020improving}, the recurrent framework~\cite{yang2020recurrent} and feature-domain compression~\cite{hu2021fvc} were employed to further improve compression efficiency. All these methods are optimized for PSNR or MS-SSIM. Veerabadran~\etal~\shortcite{veerabadran2020adversarial} proposed a GAN-based method for perceptual video compression, but has not outperformed x264 and x265 on LPIPS. Therefore, video compression aimed at perceptual quality still offers room for improvement.

\section{Proposed PLVC approach}
\subsection{Motivation}
A GAN was first introduced by Goodfellow~\etal~\shortcite{goodfellow2014generative} for image generation. It learns to generate photo-realistic images by optimizing the adversarial loss
\begin{equation} \label{GAN}
   \min_{G} \max_{D} \mathbb{E}[f(D(\bm x))] + \mathbb{E}[g(D(G(\bm y)))],
\end{equation}
where $f$ and $g$ are scalar functions, and $G$ maps the prior $\bm y$ to $p_{\bm x}$. We define $x$ as the groundtruth image and $\hat{\bm x} = G(\bm y)$ as the generated image. The discriminator $D$ learns to distinguish $\hat{\bm x}$ from $\bm x$. In the adversarial training, it pushes the distribution of generated samples $p_{\hat{\bm x}}$ to be similar to $p_{\bm x}$. As a result, $G$ learns to generate photo-realistic images. 

Later, the conditional GAN~\cite{mirza2014conditional} was proposed to generate images \emph{conditioned} on prior information. Defining the conditions as $\bm c$, the loss function can be expressed as 
\begin{equation} \label{cGAN}
   \min_{G} \max_{D} \mathbb{E}[f(D(\bm x\,|\,\bm c))] + \mathbb{E}[g(D(\hat{\bm x}\,|\,\bm c))]
\end{equation}
with $\hat{\bm x} = G(\bm y)$. The goal of employing $\bm c$ in \eqref{cGAN} is to push $G$ to generate $\hat{\bm x}\sim p_{\hat{\bm x}|\bm c}$ with the \emph{conditional} distribution tending to be the same as $p_{\bm x|\bm c}$. In other words, it learns to fool $D$ to believe that $\hat{\bm x}$ and $\bm x$ correspond to a shared prior $\bm c$ with the same conditional probability. By properly setting the condition prior $\bm c$, the conditional GAN is expected to generate frames with desired properties, \eg, rich texture, temporal consistency and coherence, \etc. This motivated us to propose a conditional GAN for perceptual video compression.

\begin{figure*}[!t]
\centering
\includegraphics[width=\linewidth]{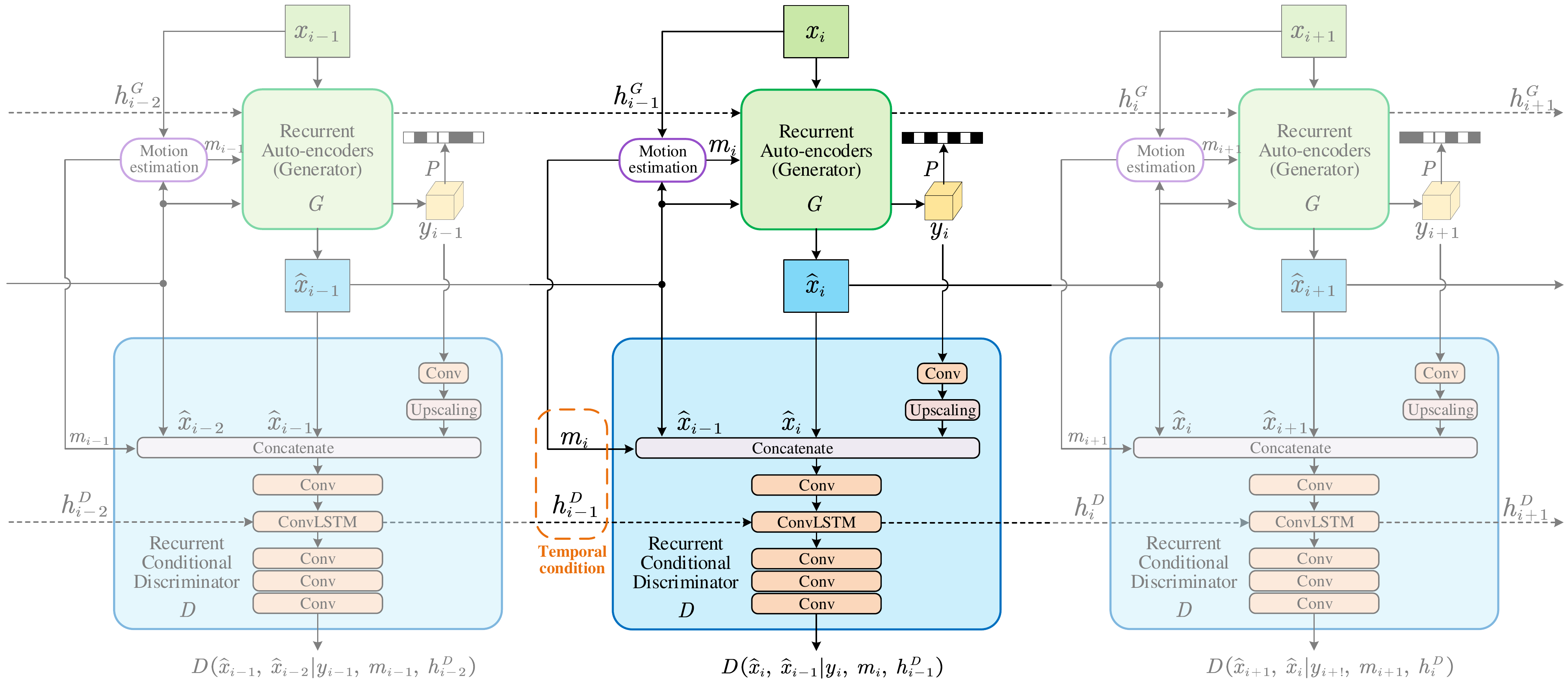}
\caption{The proposed PLVC approach with recurrent GAN, which includes a recurrent generator $G$ and a recurrent conditional discriminator $D$. The dash lines indicate the temporal information transferred through the recurrent cells in $G$ and $D$.} \label{fig:framework}
\end{figure*}

\subsection{Framework}

Fig.~\ref{fig:framework} shows the framework of the proposed PLVC approach with recurrent conditional GAN. We define the raw and compressed frames as $\{\bm x_i\}_{i=1}^T$ and $\{\hat{\bm x}_i\}_{i=1}^T$, resp. In our PLVC approach, we first estimate the motion between the current frame $\bm x_i$ and its previous compressed frame $\hat{\bm x}_{i-1}$ by the pyramid optical flow network~\cite{ranjan2017optical}. The motion is denoted as $\bm m_i$, which is to be used in the recurrent generator ($G$) for compressing $\bm x_i$ and also to serve as one of the temporal conditions in the recurrent conditional discriminator ($D$). The architectures of $G$ and $D$ in our PLVC approach are introduced next.

\textbf{Recurrent generator $G$.} The recurrent generator $G$ plays the role as a compression network. That is, given the reference frame $\hat{\bm x}_{i-1}$ and the motion $\bm m_i$, the target frame $\bm x_i$ can be compressed by an auto-encoder-based DNN, which compresses $\bm x_i$ to a latent representation $\bm y_i$ and outputs the compressed frame $\hat{\bm x}_{i}$. In our PLVC approach, we employ the recurrent compression model RLVC~\cite{yang2020recurrent} as the backbone of $G$, which contains the recurrent auto-encoders to compress the motion and the residual, and the recurrent probability model $P$ for the arithmetic coding of $\bm y_i$. The detailed architecture of $G$ is shown in the \textit{Supplementary Material}\footnote{\url{https://arxiv.org/abs/2109.03082}}. 

\textbf{Recurrent conditional discriminator $D$.} The architecture of the proposed recurrent conditional discriminator $D$ is shown in Fig.~\ref{fig:framework}.
First, we follow \cite{mentzer2020high} to feed $\bm y_i$ as the spatial condition to $D$. This way, $D$ learns to distinguish the groundtruth and compressed frames based on the shared spatial feature $\bm y_i$, and therefore pushes $G$ to generate $\hat{\bm x}_i$ with spatial features similar to $\bm x_i$. 

More importantly, the temporal coherence is essential for visual quality. We insert a ConvLSTM layer in $D$ to recurrently transfer the temporal information along time steps. The hidden state $\bm h^D_i$ can be seen as a long-term temporal condition fed to $D$, facilitating $D$ to recurrently discriminate raw and compressed video taking temporal coherence into consideration. 
Furthermore, it is necessary to consider the temporal fidelity in video compression, \ie, we expect the motion between compressed frames to be consistent with that between raw frames. For example, the ball moves along the same path in the videos before and after compression. Hence, we propose $D$ to take as inputs the frame pairs $(\bm x_i, \bm x_{i-1})$ and $(\hat{\bm x}_i, \hat{\bm x}_{i-1})$ and make the judgement based on the same motion vectors $\bm m_i$ as the short-term temporal condition. Without the $\bm m_i$-condition, $G$ may learn to generate photo-realistic $(\hat{\bm x}_i, \hat{\bm x}_{i-1})$ but with incorrect motion between the frame pair. This leads to poor temporal fidelity when compared to the groundtruth video. This motivates us to include the temporal condition $\bm m_i$ as an input to the discriminator. 

In summary, we propose to condition the discriminator $D$ on spatial features $\bm y_i$, short-term temporal features $\bm m_i$ and long-term temporal features $\bm h^D_{i-1}$. Given these conditions, we have $\bm c = [\bm y_i, \bm m_i, \bm h^D_{i-1}]$ in \eqref{cGAN} in our PLVC approach. The output of $D$ can be formulated as $D(\bm x_i, \bm x_{i-1}\,|\,\bm y_i, \bm m_i, \bm h^D_{i-1})$ and $D(\hat{\bm x}_i, \hat{\bm x}_{i-1}\,|\,\bm y_i, \bm m_i, \bm h^D_{i-1})$ for the raw and compressed samples, resp. When optimizing the adversarial loss on sequential frames, the compressed video $\{\hat{\bm x}_i\}_{i=1}^T$ tends to have the same spatial-temporal features as the raw video $\{\bm x_i\}_{i=1}^T$. Therefore, we achieve perceptual video compression with spatially photo-realistic as well as temporally coherent frames. Please refer to the \textit{Supplementary Material} for the detailed architecture of $D$. 

It is noteworthy that the effectiveness of $D$ is not restricted to a specific $G$, but generalizes to various compression networks. Please refer to the analyses and results in Section~\ref{gene}.

\subsection{Training strategies}\label{training}

Our PLVC model is trained on the Vimeo-90k~\cite{xue2019video} dataset. Each clip has 7 frames, in which the first frame is compressed as an I-frame, using the latest generative image compression approach~\cite{mentzer2020high}, and other frames are P-frames. To train the proposed network, we first warm up $G$ on by the rate-distortion loss 
\begin{equation} \label{L2}
    \mathcal{L}_w = \sum_{i=1}^N R(\bm y_i) + \lambda\cdot d(\hat{\bm x}_i, \bm x_i).
\end{equation}
In \eqref{L2}, $R(\cdot)$ denotes the bit-rate, and we use the Mean Square Error (MSE) as the distortion term $d$. Besides, $\lambda$ is a hyper-parameter to balance the rate and distortion terms.

\begin{table}[!t]
\centering
\caption{The hyper-parameters for training PLVC models.}\label{tab:hyper}
\small
\begin{tabular}{ccccccc}
\cmidrule[\heavyrulewidth]{1-7} 
Quality & $R_T$ (bpp) & $\lambda$ & $\alpha_1$ & $\alpha_2$ & $\lambda'$ & $\beta$ \\
 \cmidrule{1-7}
 Low & $0.025$ & $256$ & $3.0$ & $0.010$ & $100$ &  $0.1$\\ 
 Medium & $0.050$ & $512$ & $1.0$ & $0.010$ & $100$ & $0.1$\\ 
 High & $0.100$ & $1024$ & $0.3$ & $0.001$ & $100$ & $0.1$ \\ 
\cmidrule[\heavyrulewidth]{1-7}
\end{tabular}
\end{table}

\begin{figure}[!t]
\centering
\includegraphics[width=\linewidth]{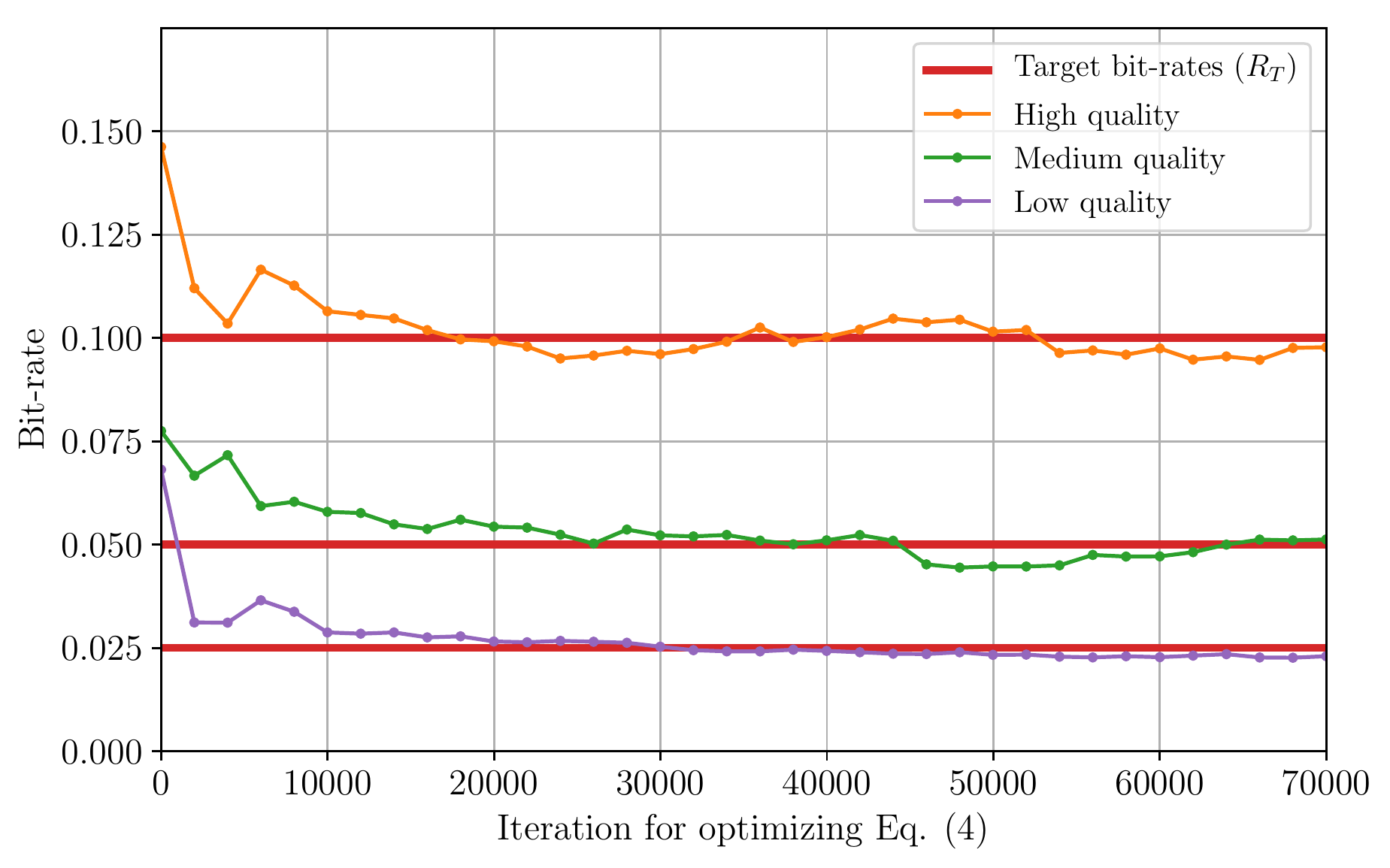}
\caption{Convergence curves of bit-rates during training.}\label{fig:train}
\end{figure}

\begin{figure*}[!t]
\centering
\subfigure{\includegraphics[width=\linewidth]{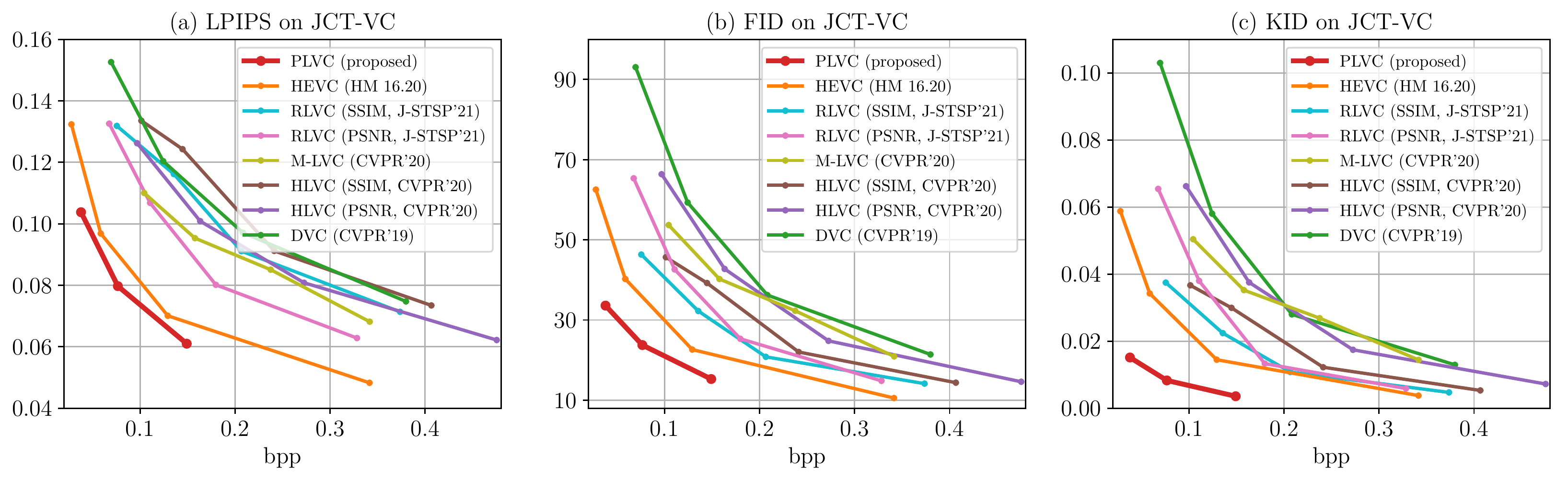}} \\
\subfigure{\includegraphics[width=\linewidth]{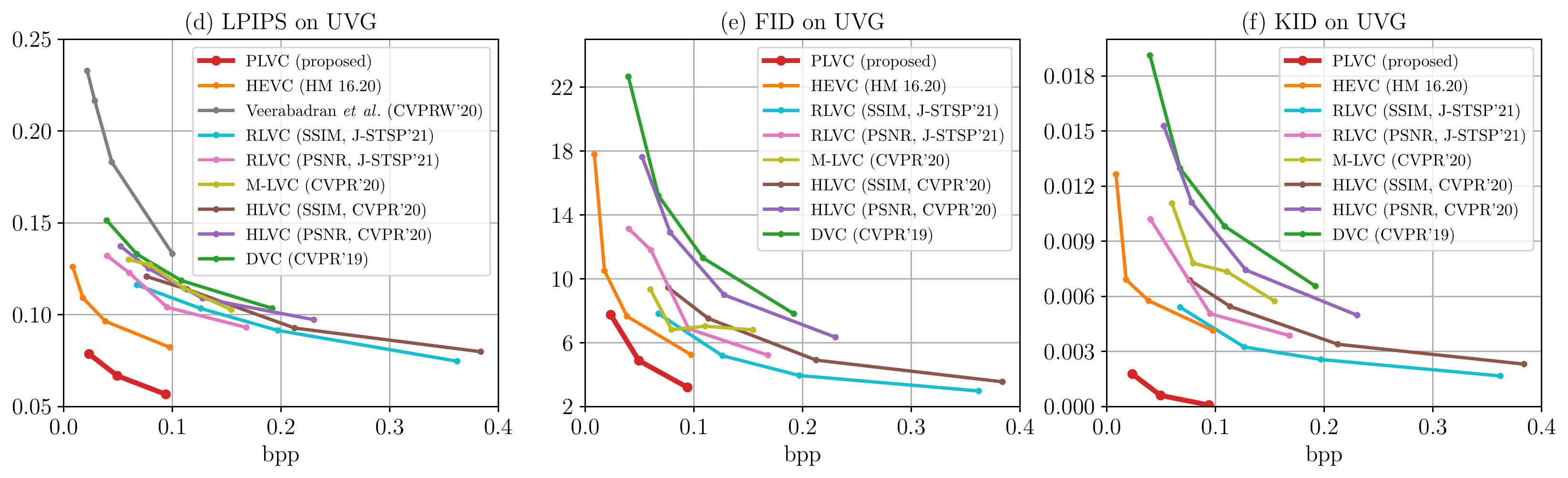}}
\caption{The numerical results on the UVG and JCT-VC datasets in terms of LPIPS, FID and KID.}\label{fig:num}
\end{figure*}

Then, we train $D$ and $G$ alternately, with the loss function combining the rate-distortion loss and the non-saturating~\cite{lucic2018gans} adversarial loss. Specifically, the loss functions are expressed as follows:
\begin{equation} \label{lossD}
\begin{aligned}
    \mathcal{L}_D &= \sum_{i=1}^N \Big(-\log\big(1-D(\hat{\bm x}_i, \hat{\bm x}_{i-1}|\bm y_i, \bm m_i, \bm h^D_{i-1})\big) 
    \\ &\quad - \log D(\bm x_i, \bm x_{i-1}\,|\,\bm y_i, \bm m_i, \bm h^D_{i-1})\Big),
    \\\mathcal{L}_{G} &= \sum_{i=1}^N \Big(\alpha\cdot R(\bm y_i) + \lambda'\cdot d(\bm{\hat{x}}, \bm x) 
    \\ &\quad - \beta\cdot \log D(\hat{\bm x}_i, \hat{\bm x}_{i-1}\,|\,\bm y_i, \bm m_i, \bm h^D_{i-1})\Big).
\end{aligned}
\end{equation}
In \eqref{lossD}, $\alpha$, $\lambda'$ and $\beta$ are hyper-parameters to control the trade-off of bit-rate, distortion and perceptual quality. We set three target bit-rates $R_T$, and set $\alpha = \alpha_1$ when $R(\bm y_i) \geq R_T$, and $\alpha = \alpha_2 \ll \alpha_1$ when $R(\bm y_i) < R_T$ to easily control bit-rates. The hyper-parameters are shown in Table~\ref{tab:hyper}.  Fig.~\ref{fig:train} illustrates the convergence curves of the bit-rates. It can be seen from Fig.~\ref{fig:train} that the bit-rates of the PLVC models converge to the target bit-rates during training, showing the effectiveness of our strategy for bit-rate control on the training set.


\section{Experiments}

\subsection{Settings}

\begin{figure}[!t]
\centering
\includegraphics[width=\linewidth]{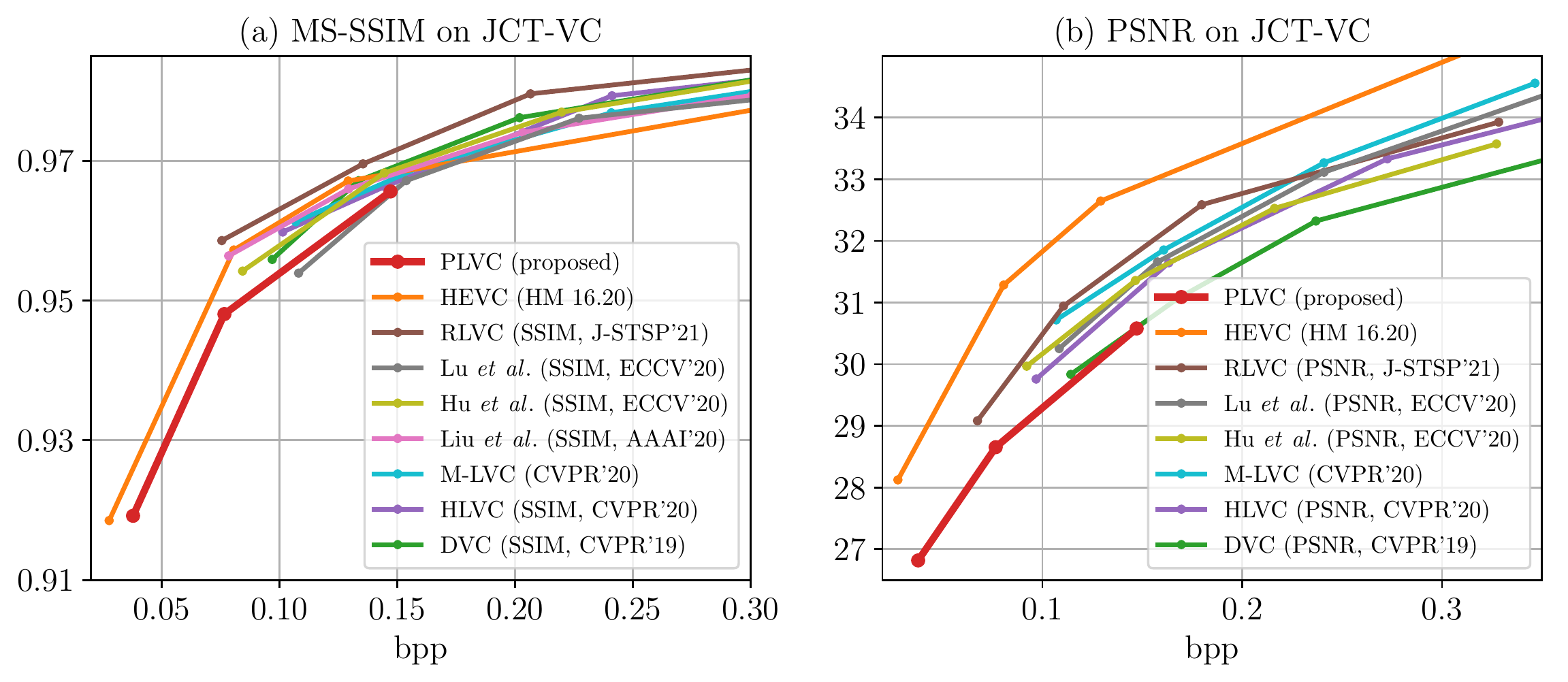}
\caption{The MS-SSIM and PSNR results on JCT-VC.}\label{fig:psnr}
\end{figure}

\begin{figure*}[!t]
\centering
\includegraphics[width=\linewidth]{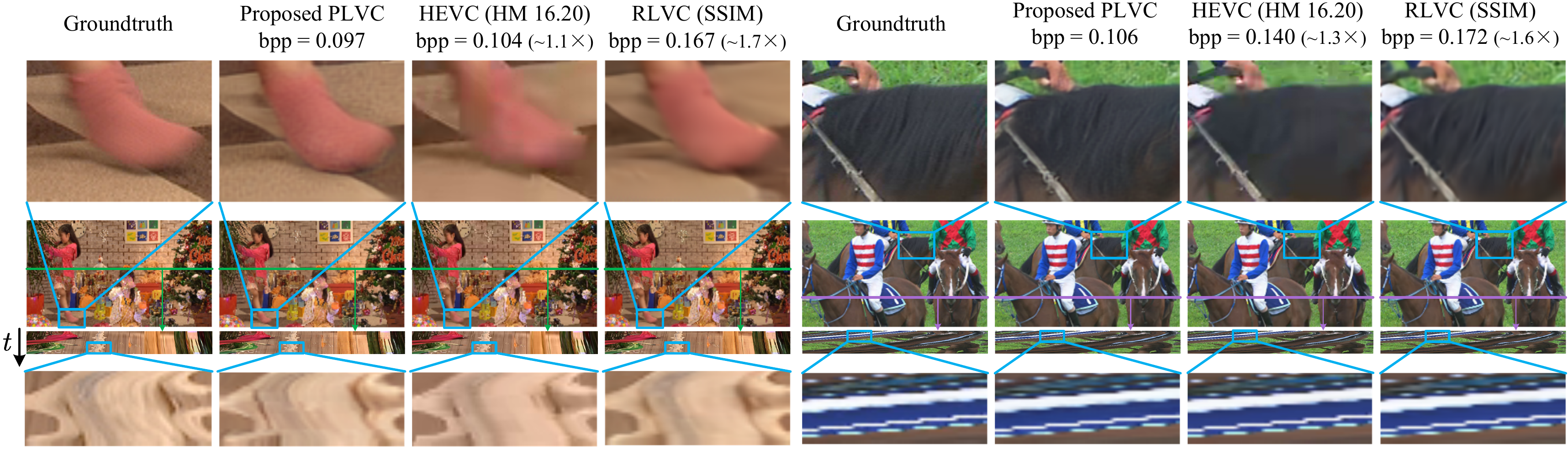}
\caption{The visual results of the proposed PLVC approach, HM 16.20 and the MS-SSIM-optimized RLVC.}
\label{fig:vis}
\end{figure*}

\begin{figure}[!t]
\centering
\includegraphics[width=\linewidth]{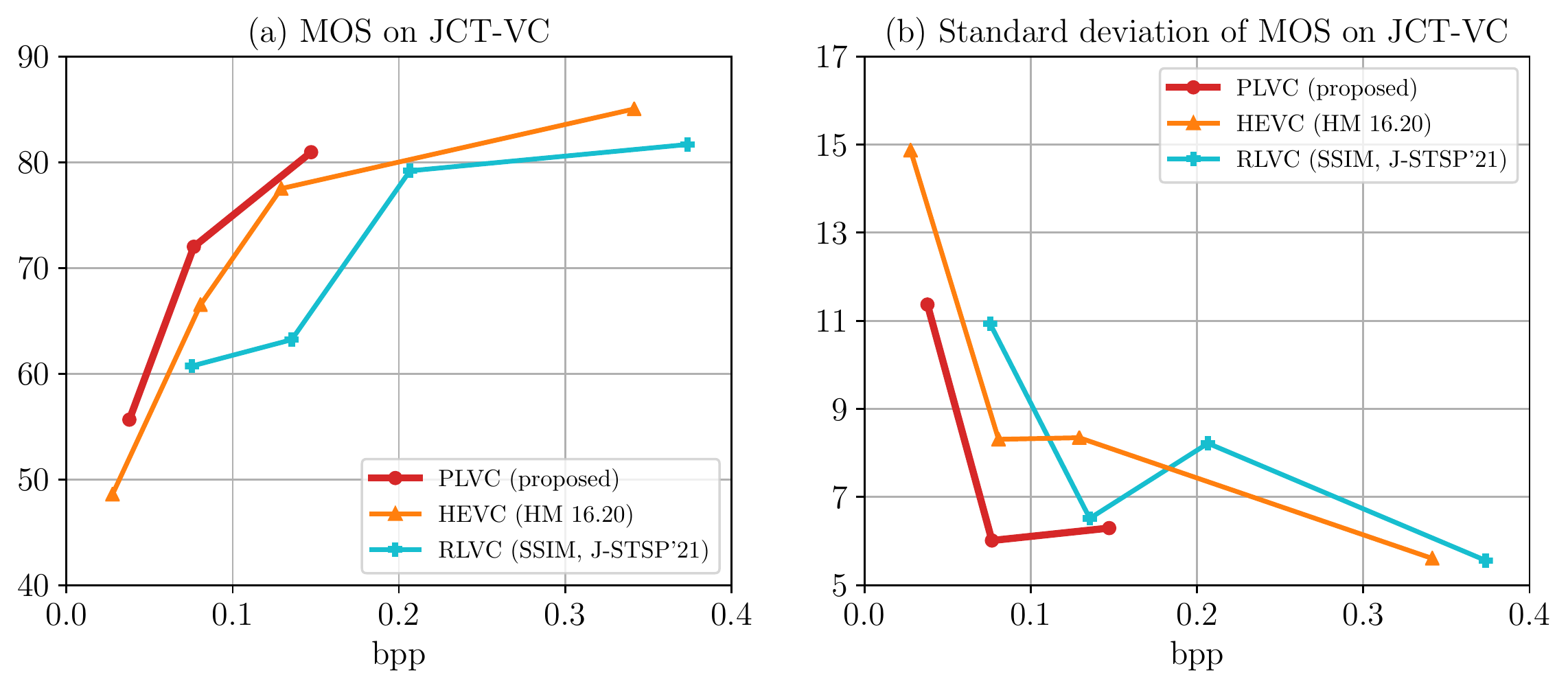}
\caption{The rate-MOS result and its standard deviation.}\label{fig:mos}
\end{figure}

We use the JCT-VC~\cite{bossen2013common} (Classes B, C and D) and the UVG~\cite{mercat2020uvg} datasets as the test sets.
On perceptual metrics, we compare with the Low-Delay P mode of the official HEVC model (HM 16.20) and the latest open-sourced\footnote{We need the codes to reproduce the videos for the user study.} learned compression approaches DVC~\cite{lu2019dvc,yang2020opendvc}\footnote{In this paper, we use OpenDVC~\cite{yang2020opendvc} to reproduce the results of DVC.}, HLVC~\cite{yang2020learning}, M-LVC~\cite{lin2020m} and RLVC~\cite{yang2020recurrent}, and the GAN-based method~\cite{veerabadran2020adversarial}. We also report the MS-SSIM and PSNR results in comparison with several existing approaches. Besides, we conduct MOS experiments as user studies to evaluate perceptual quality. There are 12 non-expert subjects participating in the MOS experiments. They are asked to rate the compressed videos with a score from 0 to 100 according to subjective quality. Higher score indicates better perceptual quality. 


\subsection{Numerical performance}

\textbf{Perceptual quality.} To numerically evaluate the perceptual quality, we calculate the Learned Perceptual Image Patch Similarity (LPIPS)~\cite{zhang2018unreasonable}, Fr\'echet Inception Distance (FID)~\cite{heusel2017gans} and Kernel Inception Distance (KID)~\cite{binkowski2018demystifying} on our PLVC and other approaches. 
In Fig.~\ref{fig:num}, we observe that the proposed PLVC approach outperforms all compared methods in terms of all three perceptual metrics. Especially, our PLVC approach reaches comparable or even better LPIPS, FID and KID values than other approaches with $2\times$ to $4\times$ the bit-rates of ours. This indicates that PLVC is able to compress video at low bit-rates but with good perceptual quality, showing the efficiency of PLVC for perceptual video compression.

\textbf{Fidelity.} Moreover, Fig.~\ref{fig:psnr} shows that the MS-SSIM of our PLVC approach is comparable with Lu~\etal~\shortcite{lu2020content}. Our PSNR result also is competitive with DVC. This shows that PLVC is able to maintain the fidelity to an acceptable degree when compressing video. Note that, although MS-SSIM is more correlated with perceptual quality than PSNR, it is still an objective metric. In Fig.~\ref{fig:psnr}-(a), all learned methods (except M-LVC which does not provide the MS-SSIM-optimized model) are specifically trained with the MS-SSIM loss, resulting in  higher MS-SSIM than the proposed PLVC. However, we have shown in Fig.~\ref{fig:num} that our PLVC approach performs best in terms of the perceptual metrics. Moreover, the user study in the following section also validates the outstanding perceptual performance of PLVC. This confirms that the perceptual metrics are better than MS-SSIM to evaluate perceptual quality.

\begin{figure}[!t]
\centering
\includegraphics[width=1\linewidth]{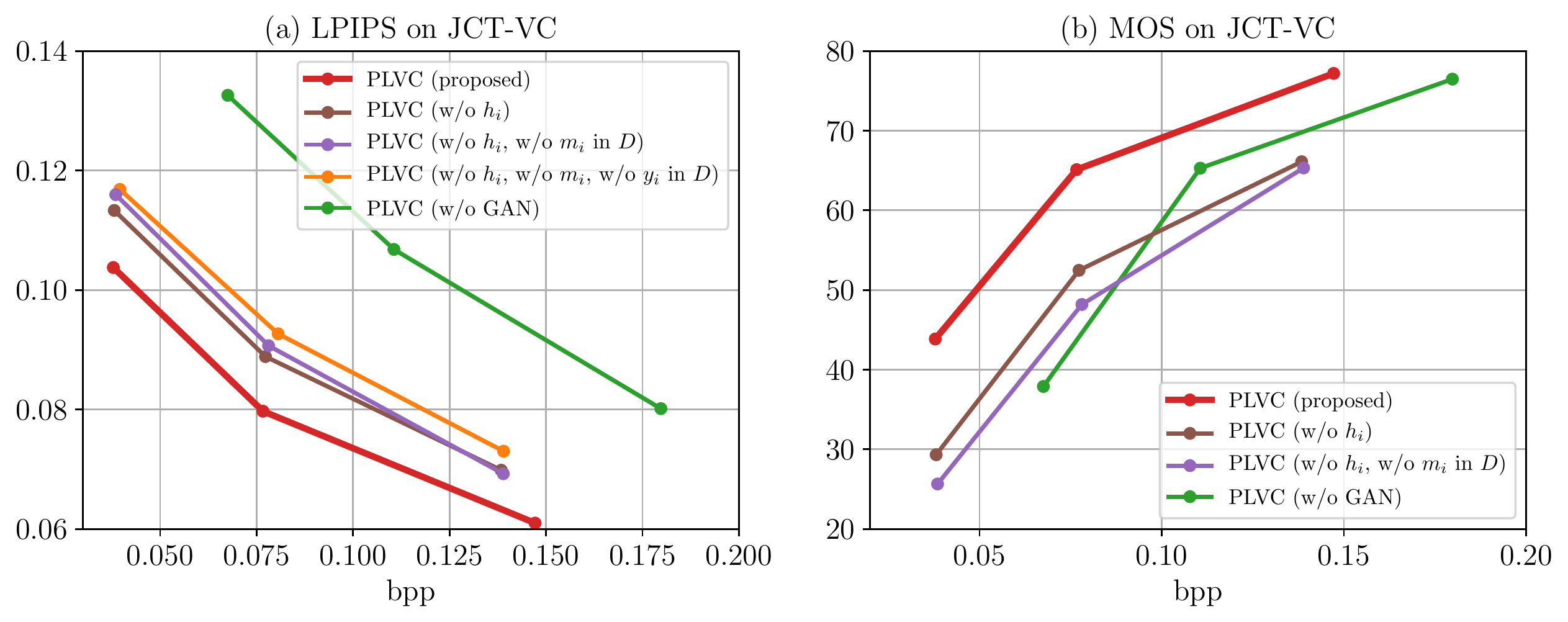}
\caption{The LPIPS and MOS results of the ablation study. This MOS experiment is independent from Fig.~\ref{fig:mos} with different subjects, so the MOS values can only be compared within each figure.} \label{fig:mos_abl}
\end{figure}

\subsection{Visual results and user study}

Fig.~\ref{fig:vis} shows the visual results of our PLVC approach in comparison with HM 16.20 and the latest learned approach RLVC~\cite{yang2020recurrent} (MS-SSIM optimized). The top of Fig.~\ref{fig:vis} illustrates spatial textures, and the bottom shows temporal profiles by vertically stacking the rows along time. It can be seen from Fig.~\ref{fig:vis} that PLVC yields richer and more photo-realistic textures at lower bit-rates than the methods compared against. Moreover, the temporal profiles indicate that PLVC maintains the temporal coherence comparable to the groundtruth. We provide more visual examples in the \textit{Supplementary Material}\footnote{\url{https://arxiv.org/abs/2109.03082}}.

\begin{figure*}[!t]
\centering
\includegraphics[width=1\linewidth]{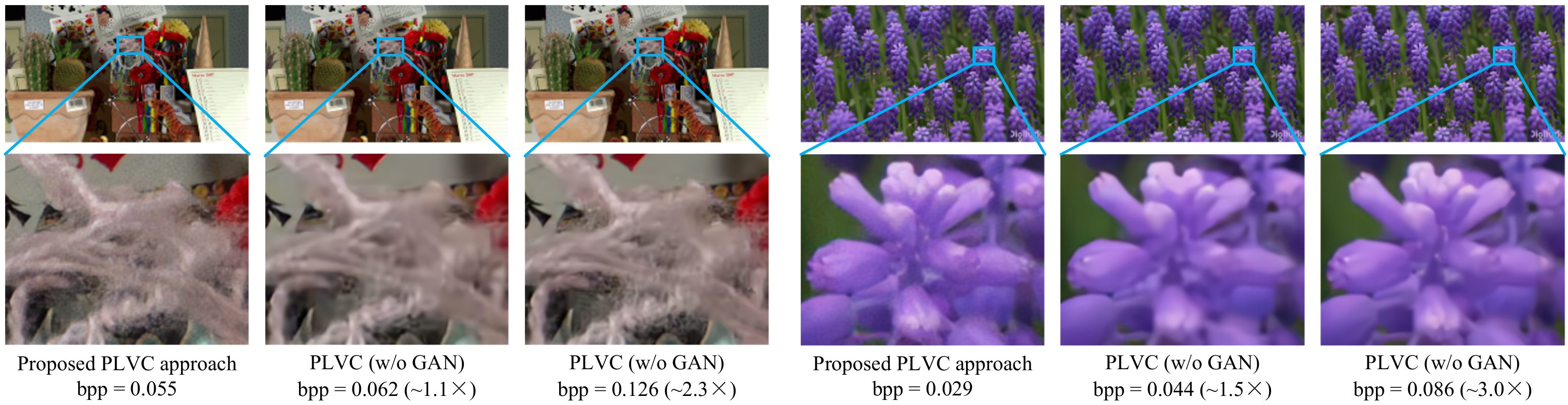}
\caption{The ablation study on PLVC with and without GAN.} \label{fig:abl_2}
\end{figure*}

\begin{figure*}[!t]
\centering
\includegraphics[width=1\linewidth]{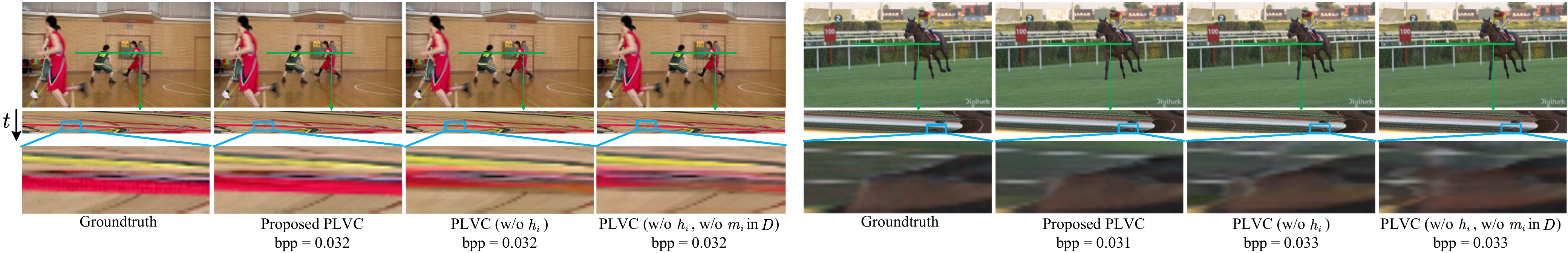}
\caption{The temporal coherence of the ablation study on the temporal conditions $\bm h_i$ and $\bm m_i$ in $D$.} \label{fig:abl_1}
\end{figure*}

\begin{figure*}[!t]
\centering
\includegraphics[width=1\linewidth]{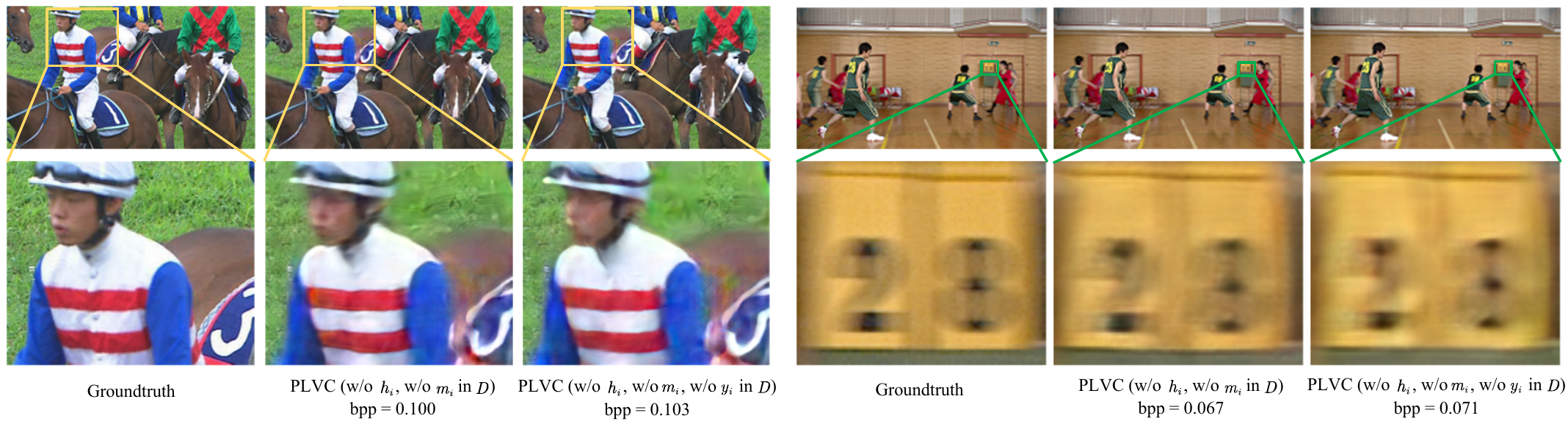}
\caption{The visual results of the models trained with and without the spatial condition $\bm y_i$ in $D$.}\label{fig:spt}
\end{figure*}

In the user study, the rate-MOS curves and the standard deviation of MOS for the JCT-VC dataset are shown in Fig.~\ref{fig:mos}. As we can see from Fig.~\ref{fig:mos}-(a), PLVC successfully outperforms the official HEVC test model HM 16.20, especially at low bit-rates, and we are significantly better than the MS-SSIM-optimized RLVC. Also, Fig.~\ref{fig:mos}-(b) shows that the standard deviation of MOS values of our PLVC approach is relatively lower than for HM 16.20 and RLVC, indicating that PLVC achieves a more stable perceptual quality. In conclusion, these results confirm that people tend to prefer the perceptual quality of PLVC over that of HM 16.20 and RLVC.

\subsection{Ablation studies}

In our ablation studies, we analyze the effectiveness of the generative adversarial network, the recurrent structure of the proposed GAN, and the spatial-temporal conditions $\bm m_i$ and $\bm y_i$. 
We provide ablation analysis in Fig.~\ref{fig:mos_abl} in terms of both LPIPS and the MOS of an ablation user study. Note that, the ablation user study is conducted independently of the user study in Fig.~\ref{fig:mos}. The subjects in these two user studies are not the same. Therefore, the MOS values can only be compared within each figure.

\textbf{Generative adversarial network.} We illustrate the results of the distortion-optimized PLVC model, denoted as PLVC (w/o GAN) in Fig.~\ref{fig:abl_2}, compared with the proposed PLVC approach. The model of PLVC (w/o GAN) is trained only until the convergence of \eqref{L2} without the adversarial loss in \eqref{lossD}. It can be seen from Fig.~\ref{fig:abl_2} that the proposed PLVC approach achieves richer textures and more photo-realistic frames than PLVC (w/o GAN), even when PLVC (w/o GAN) consumes $2\times$ to $3\times$ bit-rates. This is also verified by the LPIPS and MOS performance in Fig.~\ref{fig:mos_abl}.

\textbf{Recurrent GAN.} We first remove the recurrency in the proposed GAN (PLVC (w/o $\bm h_i$)), \ie without the hidden information transferred through the recurrent cells. As we can see from Fig.~\ref{fig:abl_1}, the temporal profile of PLVC (w/o $\bm h_i$) (third column) is obviously distorted in comparison with the proposed PLVC approach and the groundtruth. As shown in Fig.~\ref{fig:mos_abl}, the LPIPS and MOS performances of PLVC (w/o $\bm h_i$) also degrade in comparison with the proposed PLVC model. 

\textbf{Temporal condition $\bm m_i$ in $D$.} We further remove the temporal condition $\bm m_i$ from $D$ and denote it as PLVC (w/o $\bm h_i$, w/o $\bm m_i$ in $D$). As such, $D$ becomes a normal discriminator \emph{independent} along time steps. It can be seen from the right column of each example in Fig.~\ref{fig:abl_1} that the temporal coherence becomes even worse when further removing the $\bm m_i$ condition in $D$. Similar result can also be observed from the quantitative and MOS results in Fig.~\ref{fig:mos_abl}.
These results indicate that the long-term and short-term temporal conditions $\bm h_i$ and $\bm m_i$ are effective in enabling $D$ to judge raw and compressed videos according to temporal coherence, in addition to spatial texture. This way, it is able to force $G$ to generate temporally coherent and visually pleasing videos, thus resulting in good perceptual quality. 

Note that in Fig.~\ref{fig:mos_abl} the MOS values of PLVC (w/o $\bm h_i$) and PLVC (w/o $\bm h_i$, w/o $\bm m_i$ in $D$) are even lower than those of PLVC (w/o GAN) at some bit-rates. This is probably because the incoherent frames generated by PLVC without temporal conditions $\bm h_i$ and/or $\bm m_i$ severely degrade the perceptual quality, making it even worse than the distortion-optimized model. 

\begin{figure}[!t]
\centering
\includegraphics[width=.9\linewidth]{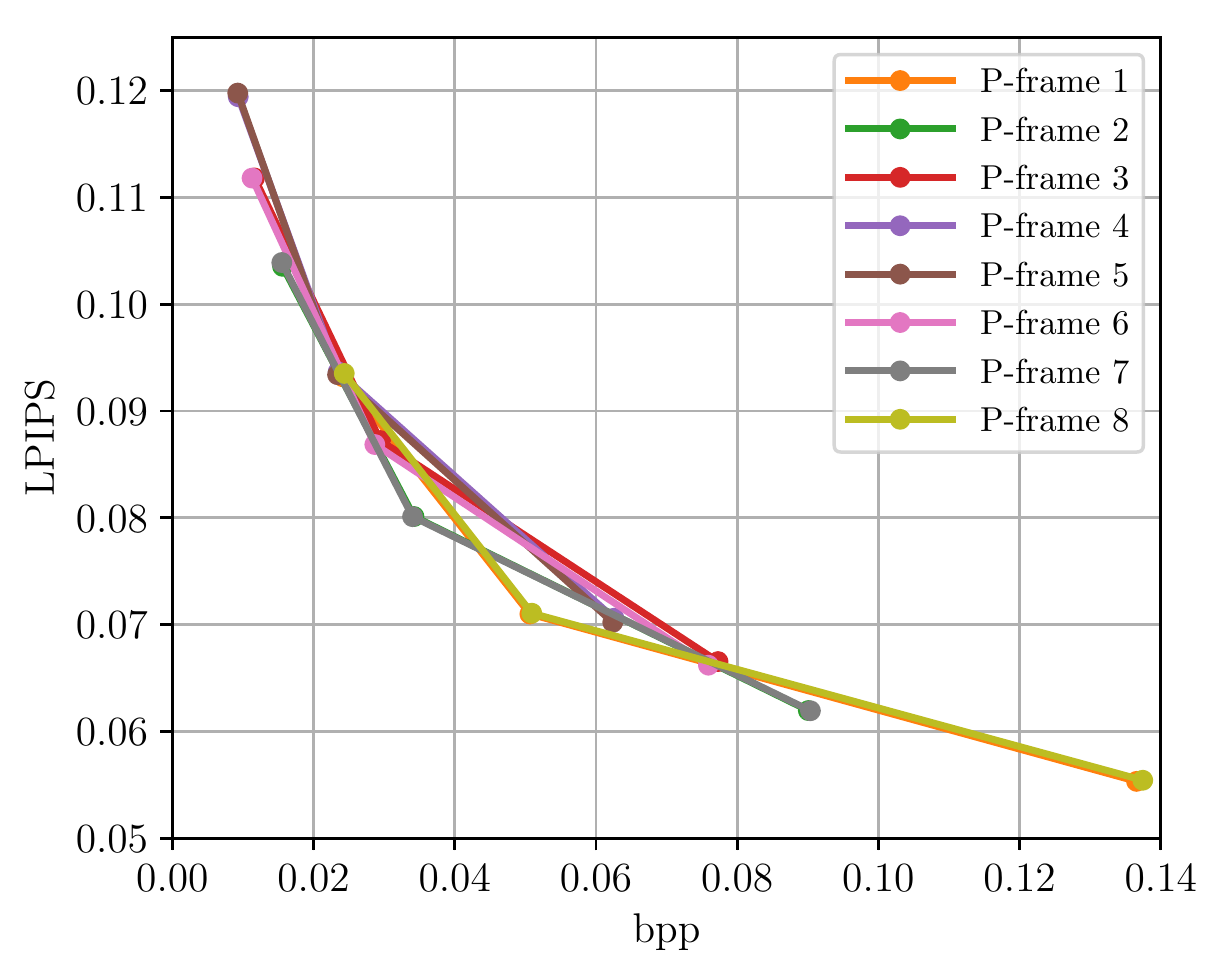}
\caption{The average performance on each P-frame in GOPs.}\label{fig:error}
\end{figure}

\textbf{Spatial condition $\bm y_i$ in $D$.} Fig.~\ref{fig:mos_abl} shows that when further removing $\bm y_i$ from the conditions of $D$ (denoted as PLVC (w/o $\bm h_i$, w/o $\bm m_i$, w/o $\bm y_i$ in $D$)), the performance in terms of LPIPS degrades further. We also show visual results in Fig.~\ref{fig:spt}.
It can be seen that, compared to PLVC (w/o $\bm h_i$, w/o $\bm m_i$ in $D$), the compressed frames generated by PLVC (w/o $\bm h_i$, w/o $\bm m_i$, w/o $\bm y_i$ in $D$) exhibit more artifacts and color shifts (the right example in Fig.~\ref{fig:spt}). This suggests that the spatial condition in $D$ is effective for pushing the generator $G$ to generate compressed frames with spatial features closer to the groundtruth. It boosts the fidelity of the compressed video and improves the visual quality.  

\subsection{GOP structure and error propagation}

In PLVC, we follow RLVC to use the bi-IPPP~\cite{yang2020recurrent} structure with the GOP size of nine frames (one I-frame and eight P-frames). Fig.~\ref{fig:error} shows the average LPIPS-bpp curve for each of the eight P-frames in GOPs. The performance is averaged among all GOPs in the JCT-VC dataset. There is no obvious error propagation in the P-frames for PLVC. The possible reasons may be two-fold. On the one hand, the bi-IPPP structure shortens the distance between I-frames and P-frames, resulting in less error propagation. On the other hand, thanks to the recurrent structure of PLVC, the hidden states contain richer temporal priors when the distance to I-frames increases. This is beneficial for compression performance, and, hence, can combat error propagation.

\begin{figure}[!t]
\centering
\includegraphics[width=\linewidth]{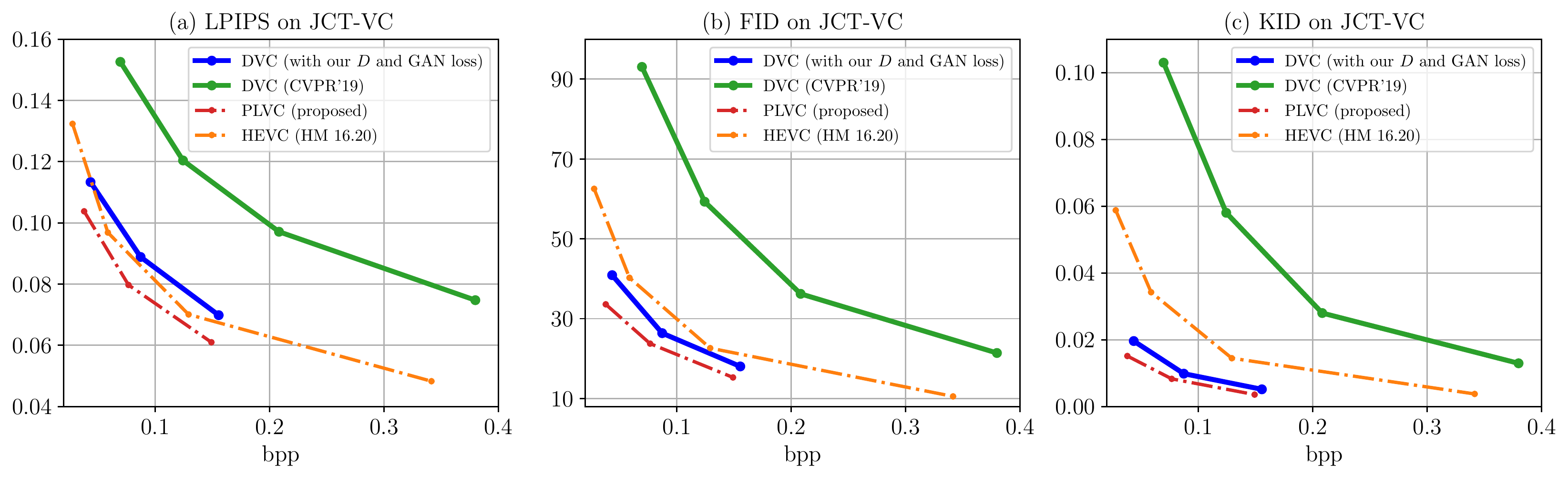}
\caption{Applying the proposed $D$ and GAN loss to DVC.}\label{fig:ge}
\end{figure}

\subsection{Generalizability of the proposed $D$}\label{gene}

Recall that in PLVC, we utilize RLVC~\cite{yang2020recurrent} as the backbone of $G$. However, the proposed recurrent conditional discriminator $D$ generalizes to various compression networks. For instance, Fig.~\ref{fig:ge} shows the results of the DVC trained with the proposed $D$ and GAN loss (\textcolor{blue}{blue} curve) compared with the original DVC (\textcolor{green}{green} curve). It can be seen that the proposed $D$ and GAN loss significantly improve the perceptual performance from the original DVC and make it outperform HM 16.20 on FID and KID. This shows that the proposed approach is not restricted to a specific compression network, but can be generalized to different backbones. It is worth mentioning that DVC trained with the proposed $D$ performs worse than PLVC, because DVC does not contain recurrent cells, while PLVC has a recurrent $G$ which works with the recurrent conditional $D$, and they together make up the proposed recurrent GAN. As a result, PLVC outperforms the previous traditional and learned video compression approaches on LPIPS, FID, KID and MOS.

\section{Conclusion}

This paper proposes a GAN-based perceptual video compression approach. Its recurrent generator learns to compress video with coherent and visually pleasing frames to fool the recurrent discriminator, which learns to assess the raw vs. compressed videos conditioned on spatial-temporal features. An adversarial loss function is designed to train the proposed model towards the trade-off among bit-rate, distortion and perceptual quality. The numerical results and user studies both validate the good perceptual performance of the proposed method, in comparison with the traditional video coding standard HM 16.20 and the learned video compression methods.

\subsection*{Acknowledgments} 
This work was supported by the ETH Z\"urich General Fund and by Humboldt Foundation.

\bibliographystyle{named}
\bibliography{ijcai22}

\clearpage

\onecolumn
\begin{center}
\begin{LARGE}
Perceptual Learned Video Compression with Recurrent Conditional GAN\\
\vspace{.5em}
-- Supplementary Material --\\
\end{LARGE}
\vspace{2em}
\begin{Large}
Ren Yang$^1$,
Radu Timofte$^{1,2}$,
Luc Van Gool$^{1,3}$

\vspace{.5em}

$^1$ETH Z\"urich, Switzerland, \\
$^2$Julius Maximilian University of W\"urzburg, Germany \quad
$^3$KU Leuven, Belgium\\

\end{Large}
\vspace{1em}
\end{center}

\begin{figure}[!b]
\centering
\includegraphics[width=\linewidth]{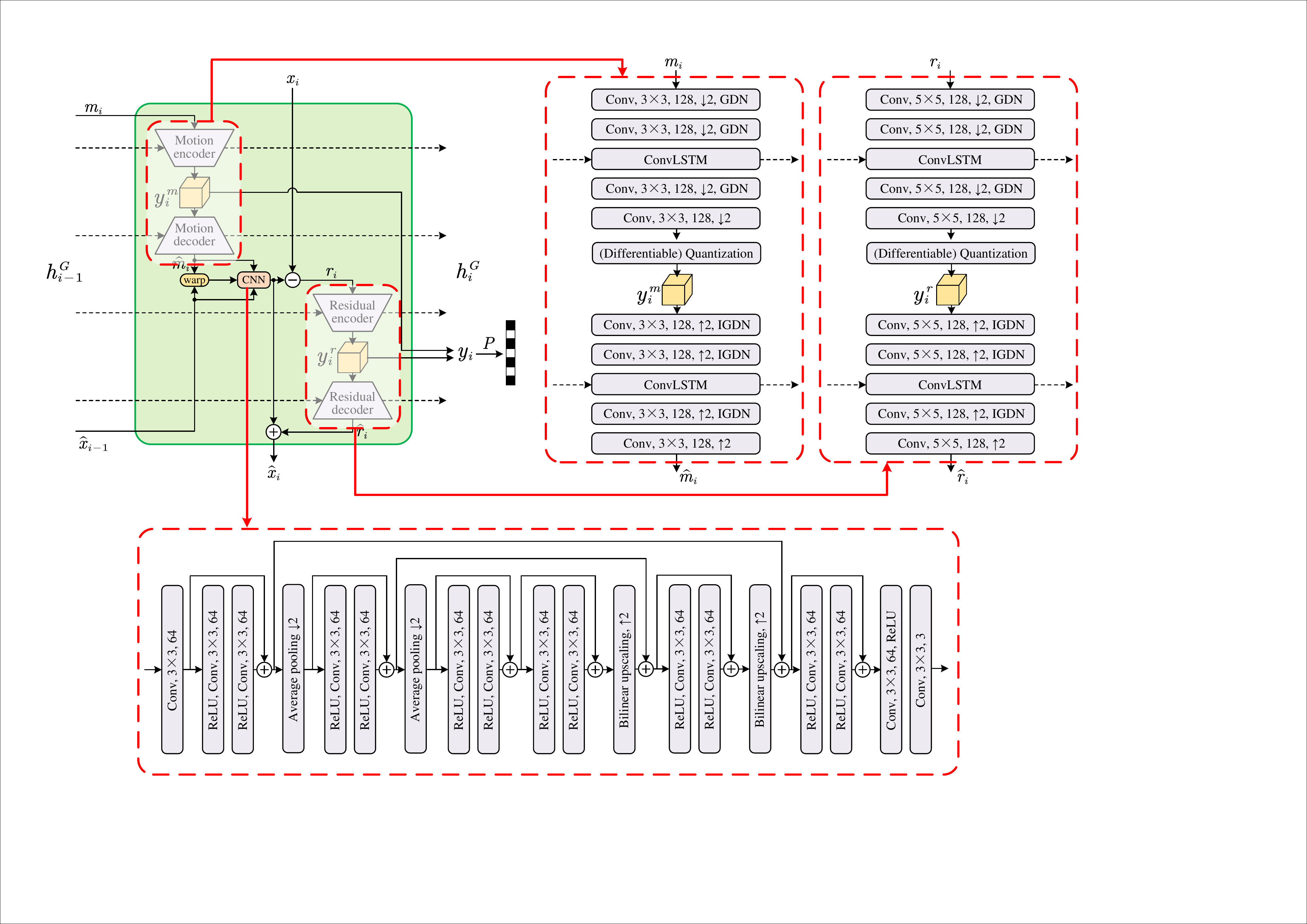}
\caption{The detailed architecture of the recurrent generator $G$.} \label{fig:G2}
\end{figure}

\section{Detailed architectures}

Fig.~\ref{fig:G2} shows the detailed architecture of the recurrent generator $G$. In Fig.~\ref{fig:G2}, the convolutional layers are denoted as ``Conv, filter size, filter number''. GDN~\cite{balle2017end} and ReLU are the activation functions. Note that the layers with ReLU before Conv indicates the pre-activation convolutional layers. $\downarrow2$ and $\uparrow2$ are $\times 2$ downscaling and upscaling, respectively. In the quantization layer, we use the differentiable quantization method proposed in~\cite{balle2017end} when training the models, and use the rounding quantization for test. The Convolutional LSTM (ConvLSTM) layers in the auto-encoders make the proposed $G$ have the recurrent structure.

\begin{figure}[!t]
\centering
\includegraphics[width=.85\linewidth]{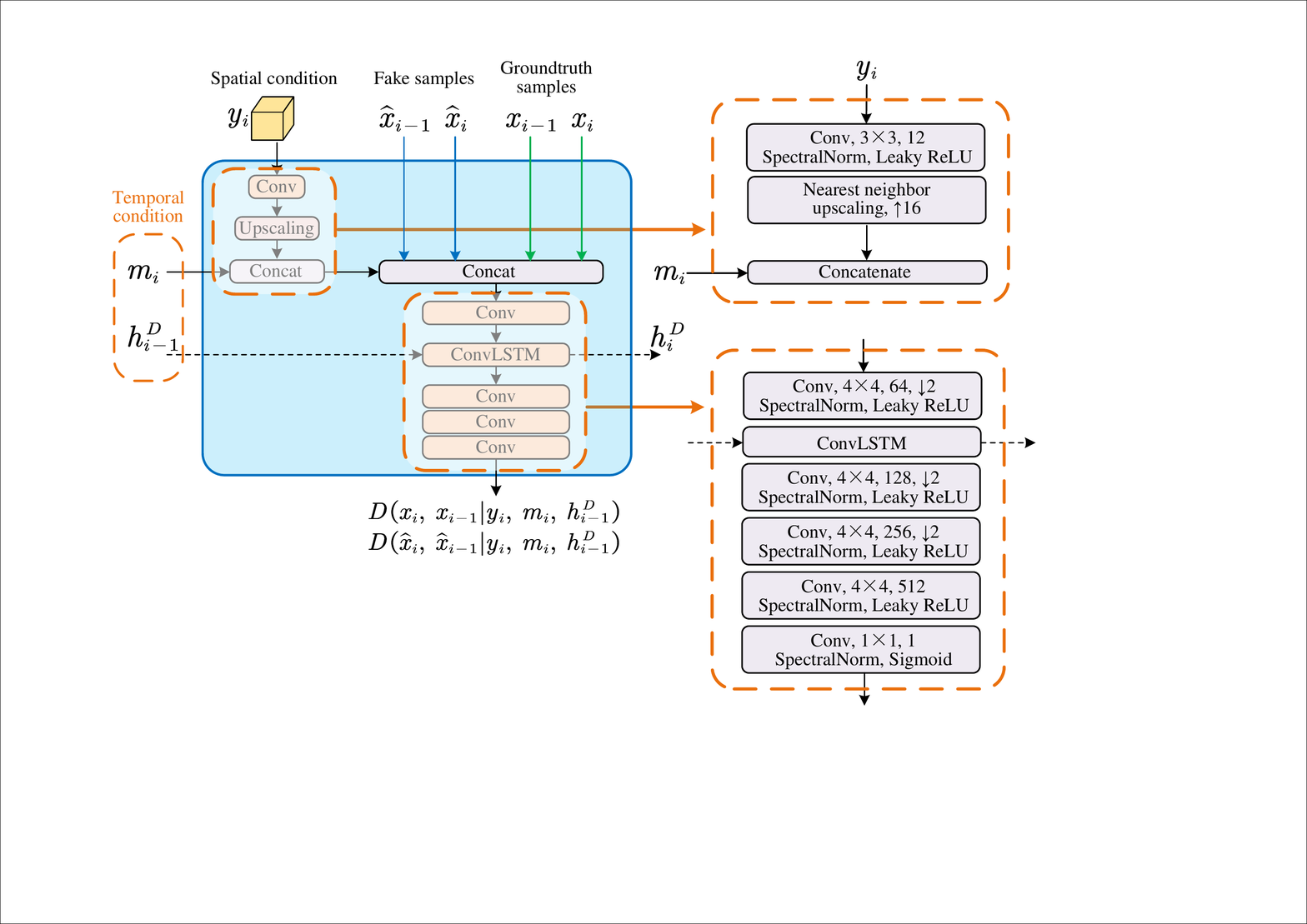}
\caption{The detailed architecture of the recurrent conditional discriminator $D$.} \label{fig:D2}
\end{figure}

The detailed architecture of the proposed recurrent conditional discriminator $D$ is illustrated in Fig.~\ref{fig:D2}. The denotations are the same as Fig.~\ref{fig:G2}. In $D$, we utilize the spectral normalization, which has been proved to be beneficial for discriminator. In the leaky ReLU, we set the leaky slope as $0.2$ for negative inputs. Finally, the sigmoid layer is applied to output the probability in the range of $[0, 1]$.

\section{More visual results}

\begin{figure}[!t]
\centering
\includegraphics[width=\linewidth]{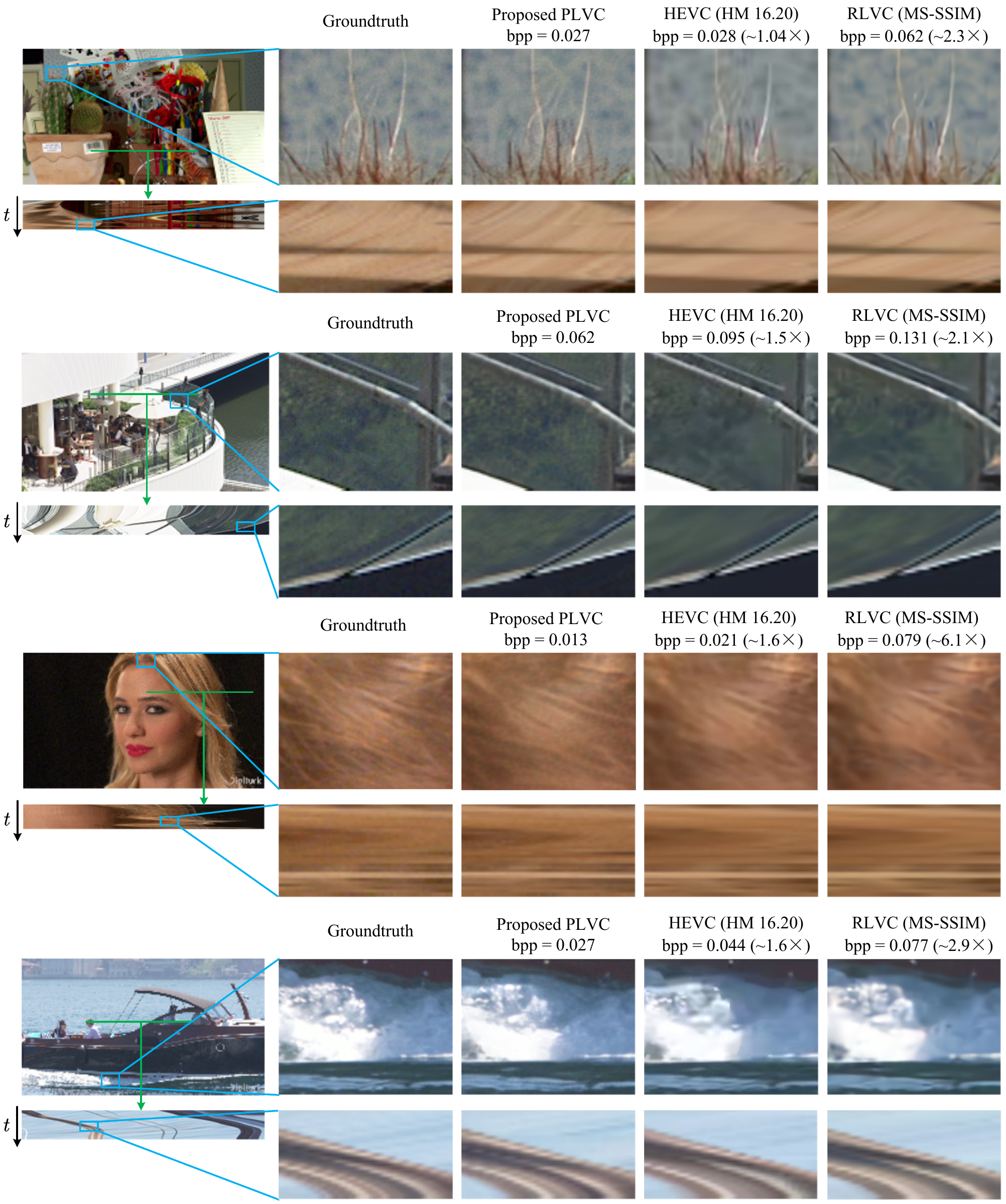}
\caption{The visual results of the proposed PLVC approach in comparison with HM 16.20 and the MS-SSIM optimized RLVC.} \label{fig:visual_2}
\end{figure}

We illustrate more visual results in Fig.~\ref{fig:visual_2} (on the \emph{last} page), which shows both spatial textures and temporal profiles of the proposed PLVC approach, HEVC (HM 16.20) and RLVC (MS-SSIM)~\cite{yang2020recurrent}, in addition to Fig.~\ref{fig:vis} in the main text. It can be seen from Fig.~\ref{fig:visual_2}, the proposed PLVC approach achieves more detailed and sharp textures than other methods, even if when HEVC consumes obviously more bit-rates and RLVC consumes more than $2\times$ bits. Besides, the temporal profiles also show that our PLVC approach has similar temporal coherence to the groundtruth, and the temporal profiles also show that we generate more photo-realistic textures than the compared methods. These results are consistent with the user study in Fig.~\ref{fig:mos} of the main text, validating the good perceptual performance of the proposed PLVC approach.

\end{document}